\documentclass[9pt,onecolumn,twoside]{pnas-new}
\templatetype{pnasresearcharticle} % Choose template 
\title{Topological Convergence of Urban Infrastructure Networks}

\author[a,1]{Christopher Klinkhamer}
\author[b]{Jonathan Zischg} 
\author[a,c]{Elisabeth Krueger}
\author[a]{Soohyun Yang}
\author[d,e]{Frank Blumensaat}
\author[f]{Christian Urich}
\author[g]{Thomas Kaeseberg}
\author[h]{Kyungrock Paik}
\author[i]{Dietrich Borchardt}
\author[g]{Julian Reyes Silva}
\author[b]{Robert Sitzenfrei}
\author[b]{Wolfgang Rauch}
\author[j]{Gavan McGrath}
\author[g]{Peter Krebs}
\author[a]{Satish Ukkusuri}
\author[a,k]{P.S.C. Rao}

\affil[a]{Lyles School of Civil Engineering, Purdue University, West Lafayette, IN, USA}
\affil[b]{Keramida Inc., Indianapolis, IN, USA}
\affil[c]{Unit of Environmental Engineering, Department for Infrastructure, University of Innsbruck, Innsbruck, Austria}
\affil[d]{Helmholtz Centre for Environmental Research, UFZ, Leipzig, Germany}
\affil[e]{Eawag, Swiss Federal Institute of Aquatic Science and Technology, Dübendorf, Switzerland}
\affil[f]{ETH Zürich, Institute of Environmental Engineering, Zürich, Switzerland}
\affil[g]{Monash Infrastructure Research Institute, Department of Civil Engineering, Monash University, VIC, Australia}
\affil[h]{Institute of Urban Water Management, Technische Universität Dresden, Dresden, Germany}
\affil[i]{School of Civil, Environmental, and Architectural Engineering, Korea University, Seoul, Korea}
\affil[j]{Helmholtz Centre for Environmental Research – UFZ, Department of Aquatic Ecosystems Analysis and Management, Magdeburg, Germany}
\affil[k]{UWA School of Agriculture and Environment, The University of Western Australia, Perth, Australia}
\affil[l]{Agronomy Department, Purdue University, West Lafayette, IN, USA}

\leadauthor{Klinkhamer} 

\authorcontributions{Please provide details of author contributions here.}
\authordeclaration{The authors declare no conflict of interest.}
\correspondingauthor{\textsuperscript{1}To whom correspondence should be addressed. E-mail: cklinkha\@purdue.edu}


\begin{abstract}
Urban infrastructure networks play a major role in providing reliable flows of multitude critical services demanded by citizens in modern cities. We analyzed here a database of 125 infrastructure networks [roads (RN); urban drainage networks (UDN); water distribution networks (WDN)] in 52 global cities, serving populations ranging from 1,000 to 9,000,000. For all infrastructure networks, the node-degree distributions, \(p(k)\), derived using undirected, dual-mapped graphs, fit Pareto distributions, \(p(k) = \alpha k^{-\gamma}\), \(k > 2\), with a mean of \(\gamma = 2.49\) and mean \(\alpha = 2.41\) [\(MSE = 4.85E-7\)]. Variance around mean \(\gamma\) reduces substantially as network size increases. Convergence of functional topology of these urban infrastructure networks suggests that their co-evolution results from similar generative mechanisms. Analysis of growing UDNs over non-concurrent 40 year periods in three cities suggests the likely generative process to be partial preferential attachment under geospatial constraints. This finding is supported by high-variance node-degree distributions as compared to that expected for a Poisson random graph. Directed  cascading failures, from UDNs to RNs, are investigated. Correlation of node-degrees between spatially co-located networks are shown to be a major factor influencing network fragmentation by node removal. Our results hold major implications for the network design and maintenance, and for resilience of urban communities relying on multiplex infrastructure networks for mobility within the city, water supply, and wastewater collection and treatment.
\end{abstract}

\dates{This manuscript was compiled on \today}
\doi{\url{www.pnas.org/cgi/doi/10.1073/pnas.XXXXXXXXXX}}

\begin{document}

\maketitle
\thispagestyle{firststyle}
\ifthenelse{\boolean{shortarticle}}{\ifthenelse{\boolean{singlecolumn}}{\abscontentformatted}{\abscontent}}{}

\dropcap{D}ifferences in multiple factors result in cities with stark contrasts in topography, climate, regulations and financial constraints that drive their designs (structure). In addition, readily apparent disparities exist in the functions of urban infrastructure networks (e.g., mobility, water supply, urban drainage, wastewater collection). Beyond structural and functional differences, infrastructure networks evolve at different rates and in diverse patterns in cities \cite{Batty1994, Lu2004,Massucci2013,Barrington2015}. These networks are comprised of components of mixed ages (because of growth and replacement) and design standards that vary substantially based on local factors (topography, climate, population, density) and management decisions (engineering design, cost, maintenance, regulations) \cite{Barrington2015,Batty2008,Batty2013}. 

Despite these dissimilarities, infrastructure networks of various types are known to follow certain universal patterns. For example, total length of road or pipes infrastructure scale sub-linearly with urban population,  consistent with economies of scale \cite{Strano2017}. Further, land-use patterns within cities as well as their physical forms are known to follow fractal geometries influenced by essential functions carried out within the city \cite{Batty2008,Batty1994}. 

\begin{table*}[t]
\centering
\caption{General Comparison of Surface (Road) and Subterranean Infrastructure Networks (Urban Drainage and Water Distribution)}
\begin{tabular}{l p{4cm} p{4cm} p{4cm}}

\textbf{Attribute} & \textbf{Road} & \textbf{Water Distribution} & \textbf{Urban Drainage} \\
\midrule

\textbf{Data Availability} & Global Availability & Restricted or Confidential & Limited Availability\\
\\
\textbf{Structure} & Highly looped; all origins to all destinations; cyclic graphs & Less looped; flows directed from one or more sources to all points; sources may be dynamic (emergency flow strategies); cyclic graphs & Typically Branching; flows directed from all points to a single outlet; acyclic graphs \\
\\
\textbf{Evolution} & Primary Driver traffic demands; mobility & \multicolumn{2}{p{8.5cm}}{Constrained by road and building placement; multi-objective optimization of costs for maximum flow efficiency, but also for conflicting interests (resilience); ultimate design chosen from pareto fronts as a tradeoff of competing objectives} \\
\\
\textbf{Optimal Design} & Full, bidirectional connectivity to all origins and destinations, full irregular grids & Loops for redundancy; valves for reliability & Similar to rivers, but less space-filling Branching/gravity driven\\
\\
\textbf{Function} & Multi origin multi destination transport & Single (or few) origin multi destination & Multi origin single (or few) destinations\\
\\
\textbf{Management / Maintenance} & Disrupts flow of traffic & \multicolumn{2}{p{8.5cm}}{Requires closing valves or diversions, disrupting water or wastewater transport; may also result in traffic disruptions on roads due to co-location} \\
\\
\textbf{Reliability} & Highly reliable, locally vulnerable to failures & Highly reliable, vulnerable to failures, impacted population may be higher than roads & Reliable, high-tolerance to failures (urban flooding); largely externalized to other networks (roads, rivers) \\
\\
\textbf{Direction of Failure Cascades} & Heavy traffic reduces lifetime of subterranean infrastructure & Bursts affect roads, leaking pipes leads to pressure losses and service disruptions & Roadways flooding; potholes, road segments collapse \\

\bottomrule
\end{tabular}
%\addtabletext{nomenclature for the TSs refers to the numbered species in the table.}
\end{table*}

Recent work shows that because of competition for space within urban areas WDN and UDNs are geospatially co-located with RNs with as much as 80\% of the length of subterranean pipe networks in European cities expected to be geospatially co-located with the RN \cite{Mair2017}. Such high degrees of geospatial co-location suggest that multiple infrastructure networks are likely to co-evolve and exhibit similar topological features, even though these networks have different layouts, and are vastly different in terms of their structure (acyclic and cyclic graphs), functions (type and directionality of flow), and tolerance for failure (frequency and consequences). Major similarities and differences in several attributes of surface (RN) and subterranean (WDN; UDN) networks are summarized in Table 1.

Recent topological analyses, based on dual representations of the water distribution (WDN) and urban drainage networks (UDN) in a large Asian city, revealed that heavy-tailed (Pareto) node-degree distributions \(p(k)\) characterize these networks \cite{Krueger2017}, consistent with findings for the topology of RNs at city, national and continental scales \cite{Strano2017,Kalapala2006}. Motivated by these findings, we assembled a database of 125 infrastructure networks of different types and sizes (RN, WDN, UDN) for 52 global cities. 

Four key questions motivating our study are: (1) how does the functional topology of infrastructure networks vary among and within cities given their diversity?; (2) is there commonality between node-degree distributions, independent of specific functional form?; (3) does a generic generative mechanism underlie the growth of these networks?; and (4) How do failures cascade across geospatially co-located infrastructure networks?

Despite multiple structural, functional, and historical differences, as we show here, we find striking convergence in the functional topologies of RN, WDN, and UDNs across our case study cities. The effects of these findings to network fragmentation are analyzed by investigating cascading failures (directed from UDNs to RNs). These results hold significant implications to network performance, stability and resilience of urban communities relying on multiple critical services.

\section*{Topology of Infrastructure Networks}

\begin{figure}%[tbhp]
\centering
\includegraphics[width=1\linewidth]{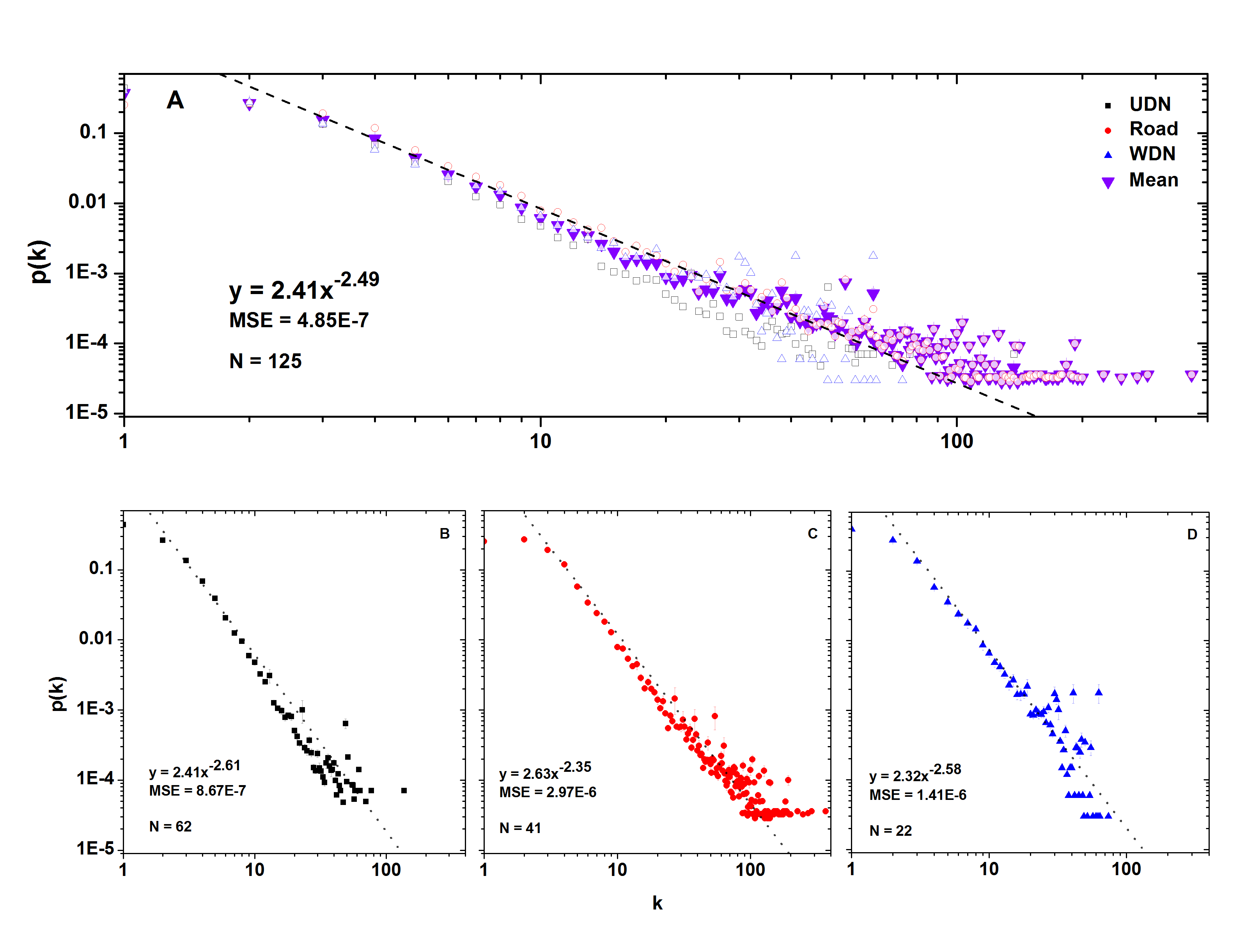}
\caption{(A) Mean p(k) (purple triangle; n = 125) of node-degree distributions (dual representation; n = 62) for all UDN (black square; n = 41) RN (red circle; n = 22), and WDNs (blue triangle (B) Mean p(k) of UDNs ; (C) Mean p(k) of RNs; (D) Mean p(k) of WDNs. Regression lines are shown for fits to Pareto probability density functions, \(p(k) = \alpha k^{-\gamma}\), \(k > 2\), with (A):  \(\alpha = 2.41\); \(\gamma = 2.49\); \(MSE = 4.85E-7\) (B): \(\alpha = 2.41\); \(\gamma = 2.61\); \(MSE = 8.67E-7\) (C): \(\alpha = 2.63\); \(\gamma = 2.35\); \(MSE = 2.97E-6\) (D): \(\alpha = 2.32\); \(\gamma = 2.58\); \(MSE = 1.41E-6\)}
\label{fig:PDF}
\end{figure}

\begin{figure}[tbhp]
\centering
\includegraphics[width=1\linewidth]{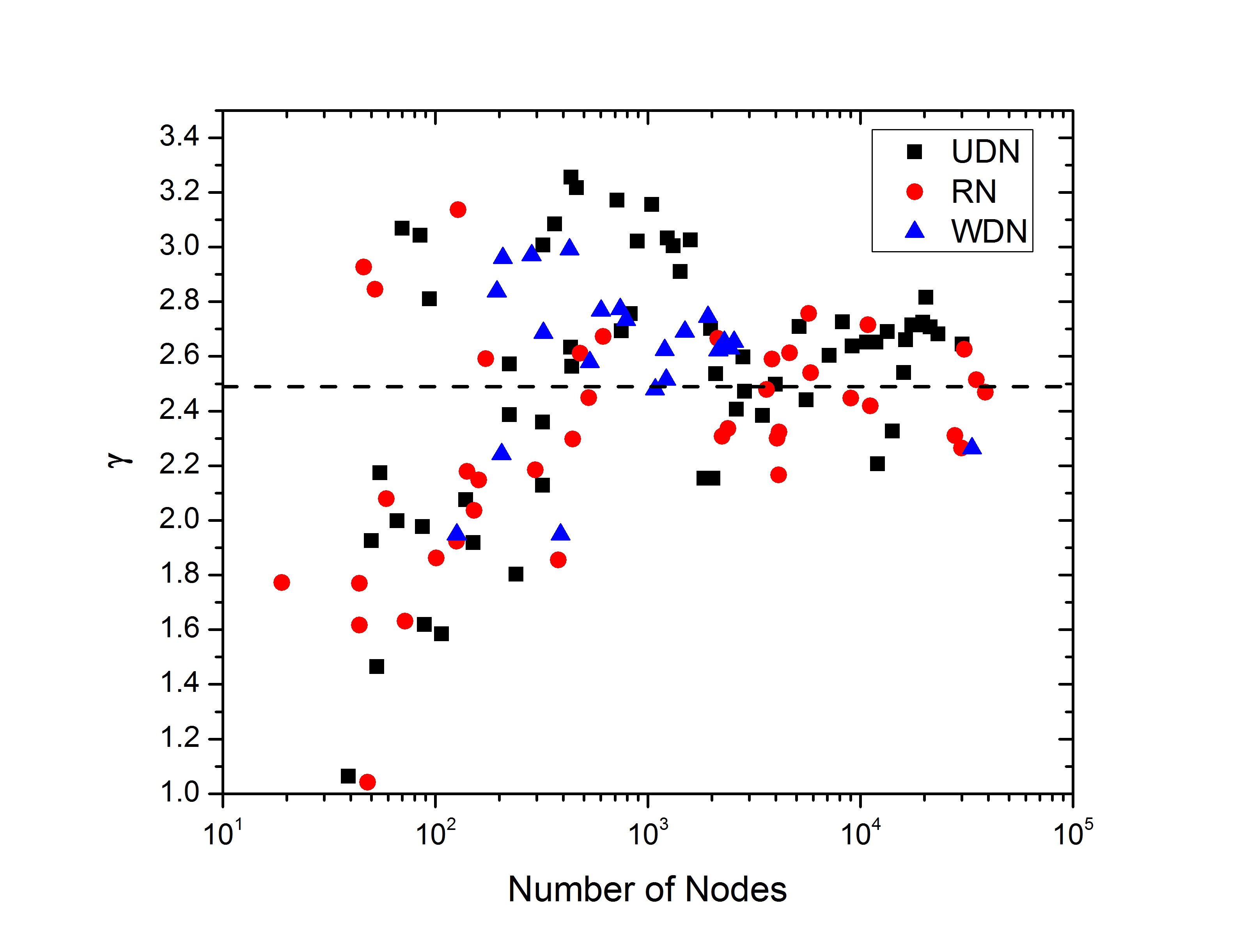}
\caption{gamma values for all UDN (black square; n = 62), RN (red circle; n = 41) and WDNs (blue triangle; n = 22). Dashed line represents mean gamma value of all networks; \(\gamma = 2.49\). \((\gamma)\) values are shown to rapidly approach the mean and decrease in variance network size increases. A reduction in variance is observed beyond a threshold network size of 2,000 nodes.}
\label{fig:AllGamma}
\end{figure} 

The structure of a network is often less important from a management standpoint, than the functions of the network. Traditional network analysis, the primal representation, of infrastructure networks, with intersections as nodes and segments as edges, reveals the structure of the network, but fails to identify key functional aspects related to the use of the networks \cite{Masucci2014}. If instead we consider the dual representation of infrastructure networks \cite{Masucci2014,Massucci2013}, where in an entire length of a road or pipe is considered a node and each intersection a link we can explore the information space of the network where functional aspects are revealed by giving importance to key attributes of network segments that influence how they are utilized (e.g., speed limit; pipe size; angle of incidence) \cite{Massucci2013,Masucci2014,Krueger2017}. Such analyses reveal universal similarities in network graphs following heavy-tailed, Pareto node-degree distributions [p(k)] \cite{Strano2017,Zhan2017,Kalapala2006,Krueger2017}. 

We begin by examining p(k) using dual representation for 125 infrastructure networks consisting of RN, UDN and WDNs in 52 global cities [see SI for details of dual representation]. We find striking consistency in their probability density functions, p(k), for all studied infrastructure networks (Figure 1) exhibit striking consistency across all three infrastructure-network types in all cities, despite distinct differences in sizes (proportional to populations served), resolution of data available, and their physical layouts. For each of the three network types, we find the mean slope \((\gamma)\) of p(k) to be in a narrow range of 2.35 to 2.6. Variability of p(k) between individual networks (Figure 2) was found to sharply decrease as network size increases, converging to values near the mean of \(\gamma = 2.49\) \([MSE <= 4.85E-7]\).

\section*{High Node-Degree Variance in Urban Infrastructure Topology}

Fitting Pareto distributions to empirical node-degree distributions is fraught with methodological challenges and controversies in interpretations regarding scale-invariance within a finite range (i.e., due to the finite size of physical networks) \cite{Clauset2009,Corral2013}. However, our goal here is not to definitively assign scale-free (or any other) distributions to these data, but instead to show that sufficient variance exists within p(k) to approximate the properties of scale-free random graphs. Multiple previous studies have shown that scale-free and other networks with highly variable node-degree distributions, identified by comparing variance to that expected of a similar random graph, are highly robust against random failure of nodes but are susceptible to the loss of high-degree nodes \cite{Albert2000,Bao2009,Song2005,Wang2008}. 

Figure 3 compares the variance in node-degree of the networks studied here to that of a Poisson random graph of equal average node degree, <k>. In all cases, the variability of the real-world infrastructure networks exceeds that of a random graph, indicating that these networks are likely to resemble “scale-free” graphs in terms of their functional topological properties, and similar failure dynamics (i.e., robustness to random failure, vulnerability to targeted attack), the latter having significant implications for urban community resilience \cite{DAgustino2014,Baxter2014,Lee2014,Baroud2014,Chan2014,DalMaso2014,DSouza2014,Lee2012,Liu2016}.

\begin{figure}[tbhp]
\centering
\includegraphics[width=.8\linewidth]{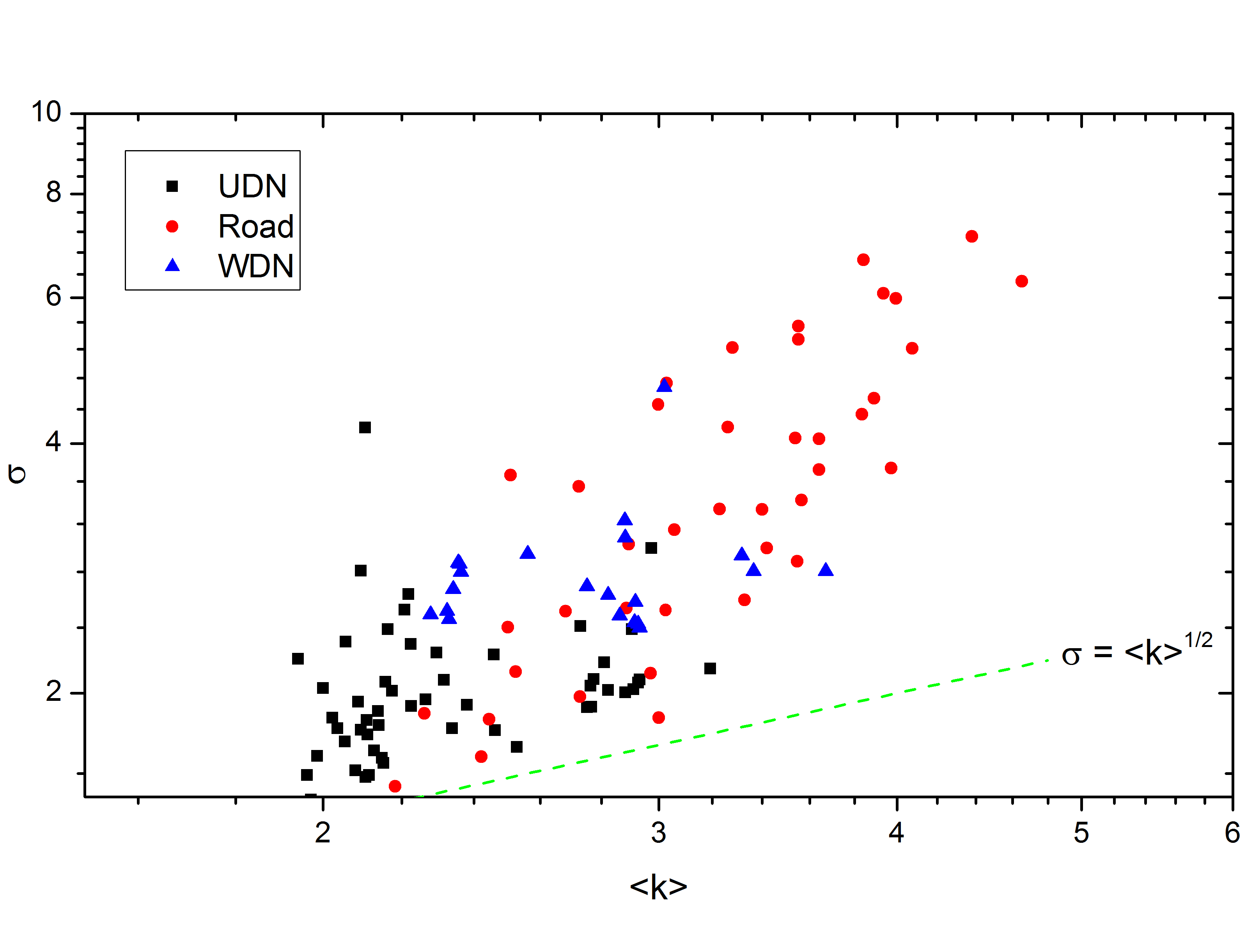}
\caption{<k> vs variance for all infrastructure networks (RN: red circle, UDN: black square, and WDN: blue triangle. The variance within the node-degree for each infrastructure network is shown to be greater than that expected of a random graph (dashed green line; \(\sigma = <k>^{1/2})\), suggesting that these networks would likely exhibit failure dynamics typical of scale-free random graphs.}
\label{fig:KvSigma}
\end{figure}

\section*{Spatially Constrained Preferential Attachment}

The convergence of functional topology of three types of urban infrastructure networks in 52 diverse cities suggests similar generative mechanisms with constraints for engineering design-optimization (cost; efficiency).  Recent studies further highlight this trend toward topological convergence, having shown that evolving networks and subnetwork components rapidly develop heavy-tailed distributions, and converge to the slope of the larger, “mature” networks \cite{Krueger2017} \cite{Yang2017}.

In network science preferential attachment is a well-known generative mechanism, involving a preference for new links added to the network to attach to existing high node-degree hubs, resulting in scale-free random graphs \cite{Barabasi1999}. This model requires that each new node entering the existing network posses complete knowledge of network connectivity in order to preferentially select an existing, high-degree hub to attach itself to. This is often not the case in real networks and the addition of new nodes confronts various constraints including, partial knowledge of connectivity, and in the case of spatial networks, topography and associated costs \cite{Carletti2015,Barthelemy2011}.Variations of the preferential attachment generative mechanism are characterized by different degrees and types of tempering\cite{Carletti2015}.

Here, we analyzed UDN growth in three different cities each over non-concurrent 40-year time lines, offering direct evidence for a variation of preferential attachment as the generative mechanism. We refer to this process as spatially constrained preferential attachment, the constraint likely being imposed by engineering and costs concerns related to the long and convoluted pipe routing that would be necessary to achieve perfect preferential attachment.

At all time-steps during the growth of these UDNs, new additions to the network are shown to be much less likely to attach to existing nodes in the 1st (lowest) quintile of node-degrees and with a preference to attach to nodes in the 2-5th quintiles (Figure 4A and B). At all time-steps, new additions to the network are most likely to attach to existing nodes in the 2\textsuperscript{nd} quintile of node-degree. This pattern of UDN growth likely results from spatial and cost constraints in the placement of new infrastructure, influencing engineers to attach new components to the nearest existing feature of sufficient capacity, resulting in a geospatially constrained preferential attachment growth mechanism.

\begin{figure}[tbhp]
\centering
\includegraphics[width=1\linewidth]{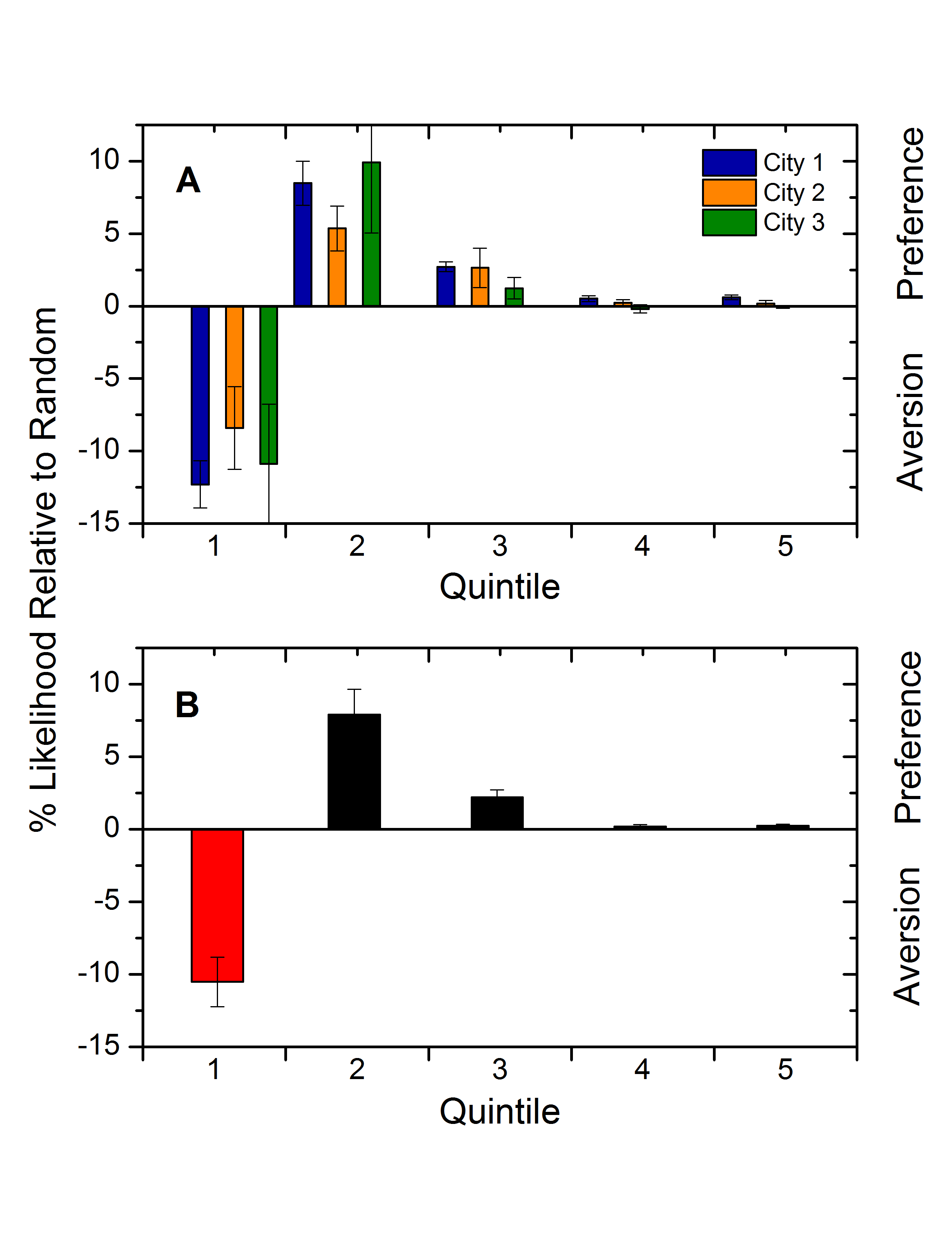}
\caption{Average attachment of of new network components for individual cities (A) and for all cities (B). In all cases there is an observed aversion for new network components to attach to existing components in the first quintile (lowest 20\%) of node-degree and a preference for attachment to second quintile (21\textsuperscript{st} - 40\textsuperscript{th}\%) and third quintile (41\textsuperscript{st} - 60\textsuperscript{th}\%) of node-degree.}
\label{fig:PrefAtt}
\end{figure}

\section*{Infrastructure Network Disruptions}

For the networks reported here we observe a mean co-location of over 60\% of the total length of the UDNs with RNs across the studied cities at a buffer distance of 15m from the road centerline (See SI). Thus, we infer that these spatially coupled infrastructure multiplex networks may be vulnerable to fragmentation due to the loss of relatively few components, particularly at smaller scales (e.g. small cities and neighborhoods). Some smaller infrastructure networks are also shown to have \(\gamma > 3\), a known threshold at which point percolation (i.e., fragmentation) transitions occur from a second-order to a first-order process \cite{Bastas2014}. As a result failures causing total fragmentation of the networks via the removal of only a small percentage of nodes are more likely in smaller networks or neighborhoods as a result of spatial variability, within subnets of a given infrastructure network. It is unlikely however that failures of this type would cascade through the extent of a large city due to neighborhood level variability. (See SI for analysis of spatial variability within infrastructure networks).

In isolation, networks with heavy-tailed p(k) (i.e. with high variance node-degree distributions) exhibit known failure dynamics wherein the underlying network is robust against fragmentation caused by random removal, but vulnerable to the targeted removal of important nodes (e.g., high node-degree; or centrality hubs) \cite{Albert2000}. However, research on coupled or multiplex networks has shown that coupling two randomly generated scale free graphs alter the normal failure dynamics of scale-free random graphs, resulting in vulnerability to failures of all types (random or targeted) \cite{DAgustino2014,Baxter2014,Lee2014,Buldyrev2010,DalMaso2014,DSouza2014,Lee2012,Liu2016}. The severity of this behavior is influenced by the orientation and connection strategies between coupled networks. However, many of these theoretical studies have been conducted on randomly generated networks that differ significantly from the empirical networks studied here. Typical assumptions include full interconnectivity, networks of equal size, and with the potential for cascades to occur in either direction between networks. All of these are assumptions that are not true of the city-scale empirical networks presented here \cite{Radicchi2014,Radicchi2015}.

Here we investigate directed cascading failures (from UDNs to RNs) for 32 cities for which data were available. Random removal of RN segments spatially co-located with UDN segments is shown to result in fragmentation of the networks (See figure 5). The speed with which the networks fragment however is variable with some networks showing greater than expected robustness (those above the dashed line and others displaying less than expected robustness (Those below the dashed line). This variance is explained by correlations between the node-degree of spatially co-located UDN and RN features, Figure 5C. A tendency for lower node-degree features to co-locate is observed, as population increases while in lower population cities, high node-degree features are more commonly located with each other leading to rapid fragmentation of the coupled networks. These results highlight the significance of separating high node-degree features when planning the city layout.

\begin{figure}%[tbhp]
\centering
\includegraphics[width=1\linewidth]{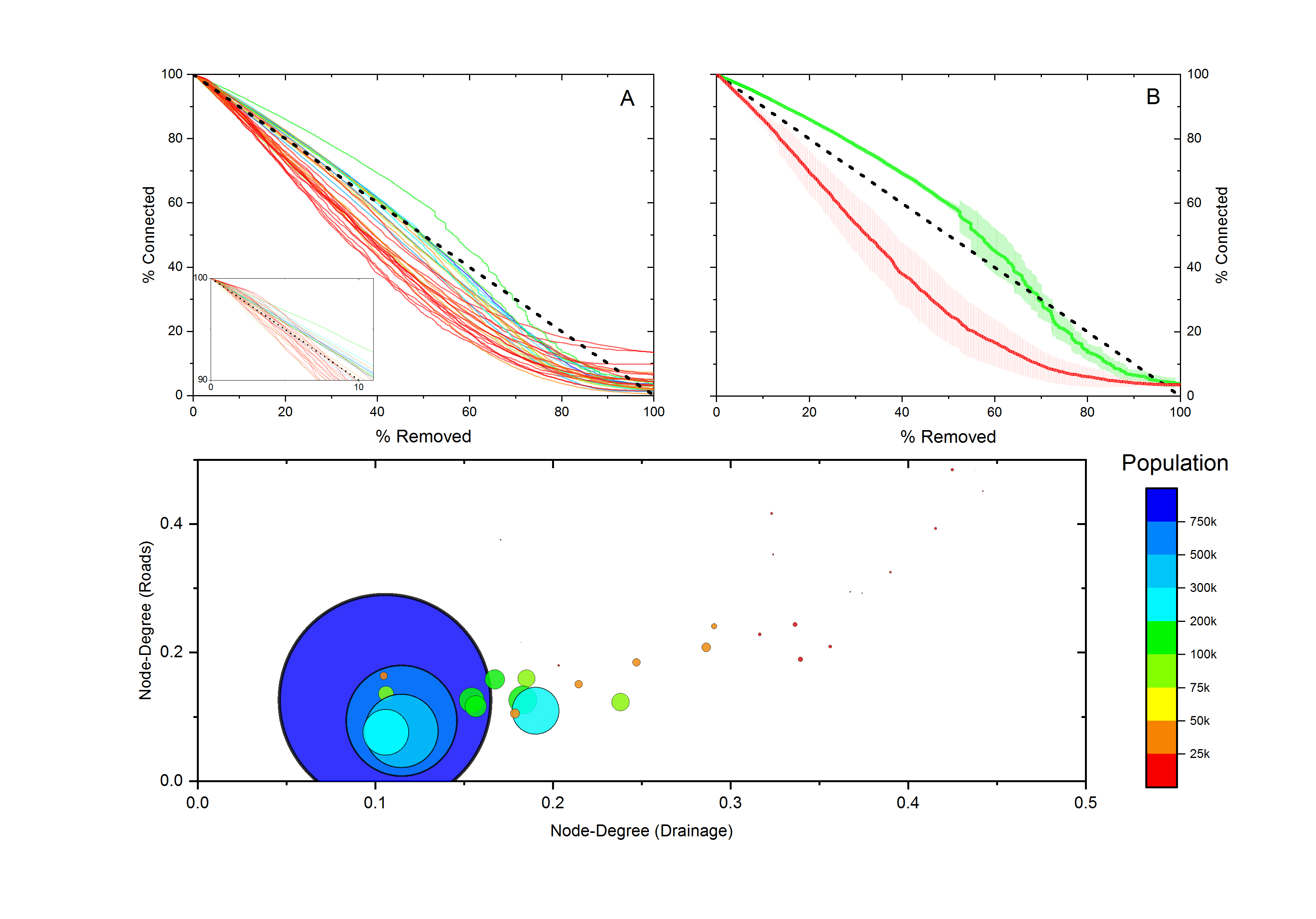}
\caption{(A) Percent connectivity of the largest connected component of all studied road networks vs percent of interconnected nodes removed. dashed-line indicates indicates 1:1 line of connectivity vs removal. networks above this line indicate greater robustness to failure than networks below the 1:1 line. Inset shows first 20\% of removal in greater detail. (B) The least (red) and most (blue) robust networks are highlighted. Shaded areas indicate standard deviation of 100 repetitions of random node removal. The more robust network retains connectivity far longer than the smaller, less robust network. (C) Correlations in node-degree between co-located drainage and road networks are shown. Smaller networks tend to have a higher ratio of co-location between high node-degree features leading to a decrease in robustness as nodes are randomly removed. Size of circles and color correspond to urban area population.}
\label{fig:Disruptions}
\end{figure}

\section*{Conclusions and Implications}

Complex networks are ubiquitous in natural, engineered and social systems, with important examples including river networks \cite{Serinaldi2016}, regional and global trade networks \cite{Xiao2017}, social networks \cite{Jackson2007}, as well as communication, mobility, and water infrastructure networks \cite{Newman2003, Newman2011, Barthelemy2011}.  Together, these interdependent, multiplex networks compose the urban fabric and provide diverse critical services in cities at multiple spatial and temporal scales. Furthermore, due to their interdependence and co-location, disruptions in one network may be able propagate to another \cite{DAgustino2014}. Therefore, characterizing the structure (topology), functions (flows), and interdependence of urban infrastructure networks has become a major topic of research in a broad range of disciplines with wide ranging applications \cite{Barthelemy2011, DAgustino2014, Batty1994, Mair2012, Rosvall2005, Guo2008}. 

Road networks (RN) are known to evolve sometimes in a decentralized growth pattern from simple grid-patterns to more complex layouts, while in other cases in the opposite direction, to increasingly more gridded patterns under centralized design and expansion \cite{Strano2012}. Quantifying the similarities in structure of globally distributed RNs has been a major focus of complex infrastructure network analysis, revealing graphs \([R(N,E)]\) that are remarkably similar despite differences in the geographical constrains, history, and design philosophies influencing the evolution of RNs over time \cite{Batty2008,Kalapala2006,Strano2012,Levinson2006}. Data for subterranean infrastructure networks, such as urban drainage networks (UDN) and water distribution networks (WDN), are not readily available due to security and confidentiality concerns. As such, these networks have received considerably less attention from a complex network analysis standpoint \cite{Mair2012,Mair2014,Strano2015,Yazdani2011,Yazdani2012}.  

Our findings suggest that a universal scaling exists for functional topology [e.g., Pareto node-degree distribution; \(p(k) = \alpha k^{-\gamma}\), \(k > 2\) for three urban infrastructure networks [RN; UDN; WDN] in 52 diverse global cities. Cities exhibit fractal geometries in terms of space-filling attributes of aboveground physical assets \cite{Batty1994,Batty2013,Strano2017,Strano2012,Strano2015}. Roads and subterranean infrastructure networks (UDN, WDN) in cities occupy the spaces between physical assets, such as buildings \cite{Mair2017}. Thus, geospatially co-located infrastructure networks exhibit comparable self-similar patterns. 

Growth of three UDNs are shown to exhibit properties of spatially constrained partial preferential attachment. Our findings also reveal these networks to highly variable node-degree distributions with heavy tailed p(k). These properties suggest that these real-world networks while robust against random failures may be vulnerable to the loss of high node-degree hubs \cite{Albert2000,Barabasi1999}. Furthermore, the observed geospatial co-location of infrastructure networks within cities introduces the possibility for cascading failures affecting multiple infrastructure networks. In an urban infrastructure context these cascading failures are likely to be predominantly directed from subterranean UDNs or WDNs to the overlying RN \cite{Strano2015,Chan2014,Kroger2014}. Examples of these types of cascades may include leaking pipes leading to soil subsidence in turn causing the collapse of road segments, or bursts of pressurized water pipes, storm events overwhelming the UDN resulting in surface flooding rendering the RN impassable \cite{Izadi2015}. Similarly, wash out of roads during large storms can also damage WDN and UDN.

Fragmentation of RNs resulting from cascading failures were shown to differ significantly from theoretical models. All cities displayed robustness to random removal of co-located features with larger cities, and those with separation of high-node-degree feature displaying increased robustness to removal.Smaller cities or subnets, may exhibit features such as near complete geospatial co-location of networks and have topological properties such as very steep NDDs \((\gamma > 3)\) that alter the nature of failure cascades across networks by shifting the percolation from a continuous, second-order transition to a discontinuous first-order transition \cite{Bastas2014}. These subnets represent opportunities for managers and city planners to address localized risks through planned maintenance or expansion by redesigning or relocating network features while maintaining city-wide performance and enhance urban community resilience.

\matmethods{Network data were obtained in the form of GIS shapefiles from a variety of sources including local governments, private companies, research institutions, and OpenStreetMap. These data were first cleaned using ESRI ArcGIS 10.1.1 to ensure network continuity. Raw shape files were cleaned and analyzed following a five-step process:

\begin{enumerate}
\item Using ArcGIS Create a geometric network from OpenStreetMap data (Snap at 0.001m, Enable complex edges)
\item Using the snapped file create a Network Dataset (Enable all vertex connectivity)
\item Export the newly created road and junction features as shapefiles
\item Run the Split Line at Points tool to split the exported road shapefile (Search radius at 0.001m)
\item Extract the graph from the split shapefile via the NetworkX Python package
\end{enumerate}

Output of this process is in the form of edge and node lists that were then analyzed in Matlab R2016b.

We estimated topological metrics for each infrastructure network by considering the dual representation, as described by Massucci et al \cite{Masucci2014}. For RNs, this process consisted of two rules that must be met for two edges (road segments) to be joined into a dual node: 1.) the angle between the two road segments must not deviate from a straight line by more than 45°, and 2.) all road segments to be joined into a dual edge must be of the same speed limit. In drainage networks, the vast majority of junctions occur at either 45 or 90°. As such the rules for joining to pipe segments into a dual node were relaxed with the only necessary criteria being that each segment to be joined be of the same diameter. All analyses in this study are based on the dual representation.

Node-degree distributions were fitted as Pareto distributions based on Maximum Likliehood Estimation, following the methods proposed by Clauset et al, and Corral and DeLuca \cite{Clauset2009,Corral2013}. We further assess the appropriateness of Pareto fits to the p(k) of the node-degree distributions using two well-accepted approaches for detecting “scale-free” properties of complex networks. First, we directly examine the variance of the node-degree distribution of the empirical pdfs and compare to that of a random network (i.e., a Poisson random graph) of identical average node-degree (<k>). Second, we identify the generative mechanism (preferential attachment) necessary to generate scale-free graphs through the analysis of the evolution of three UDNs over non-concurrent 40 year intervals at 5 year time steps

Fragmentation of the largest connected RN component was analyzed by simulating directed cascading failures (from UDNs to RNs). RN features co-located with UDN features were removed from the network one-by-one and the largest connected component was measured. Simulations were repeated 100 times to account for variability resulting from the order of node removal. All analyses were conducted using Matlab R2015B.}

\showmatmethods{} % Display the Materials and Methods section

\acknow{This research was supported by NSF Award Number 1441188 (Collaborative Research-- RIPS Type 2: Resilience Simulation for Water, Power and Road Networks).  CK was funded by the NSF grant, while EK was supported by the Helmholtz Center for Environmental Research, Leipzig, Germany, and by a Graduate Fellowship from the Purdue Climate Change Research Center. Additional financial support for the last author (PSCR) was provided by the Lee A. Reith Endowment in the Lyles School of Civil Engineering, Purdue University, and partially by the NSF grant. RS and JZ: This research is partly funded by the Austrian Research Promotion Agency (FFG) within research project ORONET (project number: 858557).}

\showacknow{} % Display the acknowledgments section

% Bibliography
\bibliography{pnas-sample}

\end{document}